\documentstyle[editedvolume]{crckapbk}
\def\sun{\hbox{$\odot$}}

\begin{opening}
\title{HOW DO THE STELLAR DISC AND THICK DISC STOP ?}

\author{A.C. Robin}
\institute{Observatoire de Besan\c{c}on, (robin@obs-besancon.fr)}
\author{M. Cr\'ez\'e}
\institute{Observatoire de Strasbourg, (creze@simbad.u-strasbg.fr)}
\author{V. Mohan}
\institute{U.P.S.O., Manora Peak, Nainital, 263129 India }

\end{opening}

\runningtitle{THE EDGE OF THE STELLAR DISC AND THICK DISC}

\begin{document}

\section{Abstract}

As part of a stellar population sampling program, a series of
photometric probes at various field sizes and depths have been obtained in a
low extinction window in the galactic anticentre direction. Very deep CCD
frames probe the most external parts of the disc,
providing strong evidence that the galactic density scale length for
the old disc population is rather
short (2.5 kpc) and drops abruptly beyond 5.5-6 kpc.

Deeper frames in the I band allow to estimate photometric distances
and confirm the position of the disc edge. A few stars are found at
larger distances. Their number is exactly what we expect if the thick
disc does not have any cutoff.
We discuss the implications for the formation and evolution of the disc,
for the star formation threshold, and for the origin of the thick disc
population.

\section{Introduction}

Efforts to sample star distributions in the galactic plane should face
both the problem of overcrowding and the complexe structure of the
absorbing layer. For these reasons the outer part of our own galactic disc is
very poorly known. The radial scale length of the density decrease is
controversial and we have nearly no indication what happens
at the end.

Three previous papers (Mohan et al. 1988, and Robin at al. 1992ab)
present the first results of a stellar population sampling program,
including a series of photometric probes at various field sizes and
depths in a low extinction window in the galactic anticentre direction.
Wide field photometry in UBV in the
magnitude range 12-17 is shown to interpret unambiguously in terms of
extinction and stellar density. The interpretation partially uses the

galactic model developed by Robin and Cr\'ez\'e (1986), Bienaym\'e et
al. (1987) and Haywood et al. (1994). The model ingredients which play a
role in the present investigation are the density law of the galactic
disc (radially exponential) and the luminosity function from Wielen et
al. (1983). Strong constraints are set on the radial structure of the
disc: the galactic disc scale length is found to be 2.5 $\pm$ 0.3 kpc.
Based on this scale length and assuming no dramatic change in the
luminosity function one can predict what should normally happen at
faintest magnitudes.

\section{Deep observations towards the anticentre}

Deep CCD observations in the UBVI bands have been obtained at the 3.6
meter CFH Telescope in a low extinction window at low latitude in the
direction of the galactic anticentre.  A detailed description of the
1987-1988 observation campaign and data analysis aspects is given
elsewhere (Robin et al. 1992a).  Observations cover four neighbouring
fields around l=179.7$\deg$ and b=2.8 $\deg$ adding up to 29 square
arcminutes. The detection limit is about magnitude 28 in V, while the
completeness limit corresponding to a photometric accuracy better than
0.1 is 25 in V and 22.5 to 24 in B depending on the frame. The 1993
observation campaign allowed to acquire V and I data complete to 25 in
both bands, allowing to have a colour index V-I down to this magnitude.

\begin{figure}
\vspace{10cm}
\caption{(V,B-V) distribution towards to the anticenter. (a) Model
prediction (dots: Disc stars at distance closer than 5.5 kpc; crosses: disc
stars at distance larger than 5.5 kpc. (b) Data sample. The thin line is just a
guide
to compare the position of stars in both diagrams. (c) V counts
(Diamonds: data with 1 sigma Poisson error
bars; dotted line: model with no disc cutoff; solid line: model with a
disc cutoff at 5.5 kpc).}
\end{figure}

\section{The edge of the old disc}

The (V, B-V) distribution of stars resulting from
this investigation is given in figure 1. The thin line is just a guide
to compare the position of stars in both diagrams.
Faint star count predictions with no cutoff in the disc deviate
strongly from the
observations at the faint end and there is a clear excess of blue
stars in the down left part of figure 1a. The bulk of disc
contributors in this magnitude range is made of disc dwarfs beyond 5.5
kpc.

All disagreements in the V, B-V diagram vanish if the stellar
disc ends abruptly at 5.5 kpc as also shown in the V star
counts in figure 1c, while the small number of remaining stars could
own to the thick disc population (see section 5 and fig. 2).

The cutoff cannot be explained by an absorbing cloud, because
external galaxies, HI column density and UBV diagrams all give a
maximum extinction of 1.2 to 1.4 magnitude.
This is also in agreement with the fact that the fields are inside
Special Area 23 selected by Kapteyn as a low extinction window.

Density distributions with scale lengths larger than the adopted 2.5
kpc would impose a still closer cutoff (at 3.5 kpc from us if h$_R$ = 3.5
kpc) while
shorter scale lengths are hardly compatible with the observations of bright
stars in the same region (Mohan et al. 1988).

\section{Where is the thick disc edge ?}

Figure 1 shows that some stars are seen at distances larger than 5.5
kpc. do these stars own to the thick disc or old disc ?

V-I colours provide a distance
indicator which may be used to grossly trace the density law,
although the V-I index used as luminosity indicator is slightly metal
sensitive. As far as we compare distances indicators computed by the
same formulae on data and on model simulations, the small error due to
metallicity variations is accounted for in the model.

We compute this apparent distance indicator using a polynomial fit of
the absolute V magnitude to the V-I index for disc stars from
the calibrations of Bessel (1991a and b) for types earlier than M4 and
Leggett (1992) for later types.

\[ M_V = 2.43 + 3.80*(V-I) - 0.0724*(V-I)^2
\]

\begin{figure}
\vspace{9cm}
\caption{Histogram of distance indicator V-M$_V$ for stars at V$<$24. Data:
solid line. Model: All stars with disc edge at 5.5 kpc : dotted line;
All stars without disc edge: dashed line; thick disc only: dotted-dashed line.}
\end{figure}

Figure~2 shows the histogram of our indicator (V-M$_V$).
The sample has been restricted to V$<$24 in order to have a more accurate
distance indicator. We see that the model with a disc edge and no
thick disc edge perfectly fit the data. The model without a disc edge
overestimate the number of stars at m-M$>$13 (r$\geq$5) by a large
factor. The number of observed stars at m-M$>$14 corresponds exactly to model
prediction for the thick disc with no cutoff.

The thick disc model results
from a global investigation of the thick disc
population based on magnitude and colour star count data
in 19 fields well distributed in longitude and latitude (Robin et
al., 1994). Such star counts turn out to tightly constrain thick disc
parameters. It has a scale
height of 760 pc, a local density 5.6\% of the disc and a scale
length of 2.5 to 3 kpc.
In the present study the use of a larger scale length for the thick
disc would have given a too large number of stars at m-M$>$14.

\section{Discussion}
The analysis of star counts towards the anticenter enable us to emphasize that
:
\begin{itemize}
\item UBV data on the magnitude range 12 to 25 from Schmidt plates and CCD
frames give and estimate a disc scale length of 2.5 $\pm$ 0.3 kpc.
\item CCD data to magnitude 23 in B and V show the existence of a
sharp cutoff in the radial distribution at
5.5 kpc from the sun (R=14 kpc).
\item I band photometry allow to go deeper in magnitude and
distance. A distance indicator is computed from V-I. It confirms the
position of the disc cutoff.
The small number of stars appearing at r>5.5 kpc is compatible
with what we expect if the thick disc has no radial cutoff.
\end{itemize}

Most external disc galaxies show a radial truncation. These cutoffs
seem to arise within 1 kpc or less (van der Kruit, 1988) and are found
at $R_{max}$ / $h_R$ = 4.5 $\pm$ 1.0. In our Galaxy star counts in
the anticentre (Robin et al. 1992a) give a radial scale length of 2.5
$\pm$ 0.3 kpc, rather short value but recently confirmed by the Fux and
Martinet (1994) kinematic determination and close to the COBE
measurement (3 kpc, Weiland et al., 1994). Together with the presently
determined cutoff, it implies a ratio $R_{max} / h$ of 5.6 $\pm$ 0.6 if
$R_{\sun}$ = 8.5 kpc in agreement with the observed ratio in external
galaxies.

Our determination of the radial extent of the old disc does not conflict
with the possibly larger extent of young stars or star forming regions
if an evolutionnary scenario like the one of Larson (1976) is
realistic (where the
star formation propagates from the centre of the Galaxy to the outer
part). In this case one expects to find only recent star
formation in the outer part of the Galaxy and no old disc
stars.

Concerning the small number of remaining stars at large distances, we
have shown that their number and distance distribution coincide
exactly with what we expect from the thick disc if it has no
cutoff. However direct measurements
of kinematics and/or metallicities of these stars would be
needed. While no accurate spectra can be obtained at
these magnitudes (V$>$23) one can get reasonably accurate proper
motions from CCD on a time baseline of ten years.

Our study implies that the thick disc does not stop like the disc in the
external part of the Galaxy. This fact confirms previous results
showing that the thick disc is
significantly different from the old thin disc.
It remains at least two scenarios compatible with
this result. The first emphasizes that the thick disc had formed
from a merger event at the beginning of the formation of the thin disc
(Quinn, Hernquist and Fullagar, 1993). These authors show that the
resulting thick disc would have a slightly larger extent than the thin
disc. The second scenario explains the
thick disc as being formed during the collapse of the gas into the
disc, stopped due to star formation (Burkert et al., 1992). In
this case as well the thick disc would extend at larger radii than the thin
disc, as observed in our data.

\acknowledgements{This research was partially supported by the Indo-French
Centre for the Promotion of Advanced Research / Centre Franco-Indien
Pour la Promotion de la Recherche Avanc\'ee.}\\\\\\


\begin{thebibliography}{}
\bibitem{}
Bessel M.S., 1991a, PASP 102,1181.
\bibitem{}
Bessel M.S., 1991b, AJ 101, 662
\bibitem{}
Bienaym\'e, O., Robin, A. C. \& Cr\'ez\'e, M. 1987,A\&A 180, 94
\bibitem{}
Burkert A., Truran J.W., Hensler G., 1992, ApJ 391, 651
\bibitem{}
Fux R. \& Martinet L., 1994, A\&A 287, L21
\bibitem{}
Haywood M., 1994, Th\`ese de doctorat, Universit\'e Paris VII.
\bibitem{}
Larson, R. B. 1976, MNRAS 176, 31
\bibitem{}
Leggett, S.K., 1992, ApJS 82, 351
\bibitem{}
Mohan, V., Bijaoui, A., Cr\'ez\'e, M. \& Robin, A. C. 1988,A\&AS 73, 85
\bibitem{}
Quinn P.J., Hernquist L. \& Fullagar D.P., 1993, ApJ 403, 74
\bibitem{}
Robin, A. C. \& Cr\'ez\'e, M. 1986,A\&A 157, 71
\bibitem{}
Robin A. C., Cr\'ez\'e M. \& Mohan V., 1992a, A\&A 253, 389
\bibitem{}
Robin A. C., Cr\'ez\'e M. \& Mohan V., 1992b, ApJ 400, L25
\bibitem{}
Robin A. C., Haywood M., Ojha D.K., Cr\'ez\'e M., Bienaym\'e O., 1994,
preprint.
\bibitem{}
van der Kruit, P. C. 1988,A\&A 192, 117
\bibitem{}
Weiland J. et al., 1994, ApJ 425, L81
\bibitem{}
Wielen, R. Jahreiss, H. \& Kr\"uger, R. 1983, In {\it The Nearby Stars and
the Stellar Luminosity Function}, I.A.U Coll. 76, p. 163 , Davis Philip and
Upgren (eds), L. Davis Press (Schenectady, NY).
\end{thebibliography}
\end{document}